\documentclass{ws-ijmpd}

\usepackage{graphics,wrapfig,times}
\usepackage{graphicx}

\usepackage{epsfig,epstopdf}
\usepackage{amsfonts}
\usepackage{mathrsfs}
\usepackage{amsmath,amssymb}
\usepackage[T1]{fontenc}

\def\CQG{\textit{Class. Quant. Grav.}\ }
\def\PR{\textit{Phys. Rev.}\ }
\def\PL{\textit{Phys. Lett.}\ }

\def\RMP{\textit{Rev. Mod. Phys.}\ }

\begin{document}

\markboth{Fei-Hung Ho \& James M. Nester}
{PGT with coupled even \& odd parity dynamic spin-0 modes
}

\title{POINCAR\'E GAUGE THEORY WITH COUPLED EVEN AND ODD PARITY DYNAMIC SPIN-0 MODES: DYNAMICAL EQUATIONS FOR ISOTROPIC BIANCHI COSMOLOGIES}

\author{Fei-Hung Ho}

\address{Department of Physics, National Central University\\
Chungli, 320, Taiwan\\
93242010@cc.ncu.edu.tw}

\author{James M. Nester}

\address{Department of Physics, National Central University\\
Chungli, 320, Taiwan\\
Graduate Institute of Astronomy, National Central University\\
Chungli 320, Taiwan\\
Center for Mathematics and Theoretical Physics,
National Central University\\
Chungli 320, Taiwan\\
nester@phy.ncu.edu.tw}

\maketitle


\begin{abstract}
We are investigating the dynamics of a new Poincar\'e gauge theory of gravity model, which has cross coupling between the spin-0$^+$ and spin-0$^-$ modes.  To this end we here consider a very appropriate situation---homogeneous-isotropic cosmologies---which is relatively simple, and yet all the modes have non-trivial dynamics which reveals physically interesting and possibly observable results.  More specifically we consider manifestly isotropic Bianchi class A cosmologies; for this case we find an effective Lagrangian and Hamiltonian for the dynamical system.  The Lagrange equations for these models lead to a set of first order equations that are compatible with those found for the  FLRW models and provide a foundation for further investigations.  Typical numerical evolution of these equations shows the expected effects of the cross parity coupling.

 \keywords{gravity, gauge theory, cosmology, parity}
\end{abstract}



\section{Introduction}

All the known fundamental physical interactions can be  formulated
in a common framework: as {\em local gauge theories}.
However the standard theory of gravity, Einstein's general relativity (GR), based on the spacetime metric, is a
rather unnatural gauge theory.
Physically (and geometrically) it is reasonable to consider gravity as a
gauge theory of the local Poincar\'e symmetry of Minkowski
spacetime.  A theory of gravity based on local spacetime geometry gauge symmetry, the quadratic Poincar\'e gauge theory of gravity (PG, aka PGT) was worked out some time ago.\cite{HHKN,Hehl80,HS80,MieE87,HHMN95,GFHF96,Blag02}
We  briefly sketch this theory, and how the search for good dynamical propagating modes led to focusing on the two scalar modes.

There is no known fundamental reason why the gravitational coupling should respect parity. With this in mind,
the general quadratic PG theory has recently seen renewed interest in including all possible odd parity couplings. The appropriate cross parity pseudoscalar coupling constants have been been incorporated, in particular, into the special case of the dynamically favored two scalar mode model to give the BHN model,\cite{BHN11} which is the most general PG model that we expect to have problem free dynamics.  Here we show how a simple effective Lagrangian can reveal in a cosmological context the dynamics of this extended model.

We are especially interested in investigating the {\em dynamics} of the PG BHN model.  This can be expected to be very clearly revealed in purely time dependent solutions, hence we considered {\em homogeneous cosmologies}.  The two dynamical connection modes that we wish to study carry spin 0$^+$ and spin 0$^-$ (and are thus referred to as {\em scalar} modes or more specifically as the {\em scalar} and {\em pseudoscalar} mode).  Consequently in a homogeneous situation they cannot pick out any spatial direction, and thus they have no interaction with spatial anisotropy, so for a study of their dynamics it is sufficient (and most simple) to confine our attention to {\em isotropic} models.  Here, we present an extension of the results briefly reported in Ref.~\refcite{iard10}.  In particular for the PG BHN model we find an effective Lagrangian and Hamiltonian as well as a system of first order dynamical equations for Bianchi class A isotropic homogeneous cosmological models and present some sample evolution which shows the effect of the cross parity coupling.

\section{The Poincar\'e gauge theory}

In the Poincar\'e gauge theory of gravity,\cite{HHKN,Hehl80,HS80,MieE87,HHMN95,GFHF96,Blag02}
the two sets of local gauge potentials are, for ``translations'',
the orthonormal co-frame $\vartheta^\alpha=e^\alpha{}_i {\rm d}x^i$,
where the metric is
$g=-\vartheta^0\otimes\vartheta^0+\delta_{ab}\vartheta^a\otimes\vartheta^b$,
and, for ``rotations'', the metric-compatible (Lorentz Lie-algebra
valued) connection 1-forms
$\Gamma^{\alpha\beta}=\Gamma^{[\alpha\beta]}{}_i {\rm d}x^i$. The
associated field strengths are the torsion and curvature 2-forms
\begin{eqnarray}
T^\alpha&:=&{\rm d}\vartheta^\alpha+\Gamma^\alpha{}_\beta\wedge
\vartheta^\beta=\frac12 T^\alpha{}_{\mu\nu}\vartheta^\mu\wedge\vartheta^\nu,
\label{torsion}\\
R^{\alpha\beta}&:=&{\rm d}\Gamma^{\alpha\beta}
+\Gamma^\alpha{}_\gamma\wedge\Gamma^{\gamma\beta}=
\frac12R^{\alpha\beta}{}_{\mu\nu}\vartheta^\mu\wedge\vartheta^\nu,
\label{curvature}
\end{eqnarray}
which satisfy the respective Bianchi identities:
\begin{equation}
DT^\alpha\equiv R^\alpha{}_\beta\wedge \vartheta^\beta,\qquad
DR^\alpha{}_\beta\equiv0.\label{bianchi}
\end{equation}

Turning from kinematics to dynamics, the PG Lagrangian density is generally taken
to have the standard quadratic Yang-Mills form, which leads to quasi-linear second order equations for the gauge potentials. Qualitatively,
\begin{equation}
\quad{\mathscr L}[\vartheta,\Gamma]\sim \kappa^{-1}[\Lambda+\hbox{curvature}+\hbox{torsion}^2]+\varrho^{-1}\hbox{curvature}^2,
\end{equation}
where $\Lambda$ is the cosmological constant, $\kappa=8\pi G/c^4$, and $\varrho^{-1}$ has the dimensions of action.
The field equations, including source terms, obtained by variation w.r.t. the two gauge potentials have the respective general forms
\begin{eqnarray}
\Lambda+\hbox{curvature}+D\hbox{ torsion}+\hbox{torsion}^2+\hbox{curvature}^2
&\sim&\hbox{energy-momentum density},\\
\hbox{torsion}+D\hbox{ curvature}&\sim&\hbox{spin density}.
\end{eqnarray}
From these two equations, with the aid of the Bianchi
identities (\ref{bianchi}), one can obtain, respectively,
the conservation of source energy-momentum and angular momentum
statements.

Earlier investigations generally considered  models with even parity terms; the models had 10 dimensionless
scalar coupling constants.
The recent BHN investigation\cite{BHN11} systematically considered
all the possible odd parity Lagrangian terms, introducing 7 new pseudoscalar coupling constants.
Not all of these coupling constants are physically independent, since there are 3 topological invariants: the (odd parity) Nieh-Yan\cite{NY} identity
$d(\vartheta^\alpha\wedge T_\alpha)\equiv T^\alpha\wedge T_\alpha+R_{\alpha\beta}\wedge\vartheta^{\alpha\beta}$,
the (even parity) Euler 4-form $R^{\alpha\beta}\wedge R^{\gamma\delta}\eta_{\alpha\beta\gamma\delta}$,
and the (odd parity) Chern-Symons 4-form $R^\alpha{}_\beta\wedge R^\beta{}_\alpha$.
For detailed discussions of the BHN Lagrangian and the topological boundary terms see Ref.~\refcite{BHN11} and the new work Ref.~\refcite{BH11}.


Early PG  investigations (especially Refs.~\refcite{HS80,SN80}) of the linearized
theory identified six possible dynamic connection modes; they
carry spin-$2^{\pm}$, spin-$1^{\pm}$, spin-$0^{\pm}$. A good dynamic
mode should transport positive energy and should not propagate
outside the forward null cone. The linearized investigations found
that at most three modes can be simultaneously dynamic; all the
acceptable cases were tabulated; many combinations of three modes
are satisfactory to linear order. Complementing this, the
Hamiltonian analysis revealed the related constraints.\cite{BMNI83}
Then detailed investigations of the Hamiltonian and propagation\cite{CNY98,HNZ96,yo-nester-99,yo-nester-02} concluded that effects
due to nonlinearities in the constraints could be expected to render
all of these cases physically unacceptable except for the two
``scalar modes'', carrying spin-$0^+$ and spin-$0^-$.

In order to further investigate the dynamical possibilities of these PG scalar modes,  Friedmann-Lema\^{i}tre-Robinson-Walker (FLRW)
 cosmological models were considered.  Using a $k=0$ model it was found that the $0^+$ mode naturally couples to the acceleration of the universe and could account for the present day observations;\cite{YN07,SNY08} this model was then extended to include the $0^-$ mode.\cite{JCAP09}

After developing the general odd parity PG theory, in BHN\cite{BHN11} the two scalar torsion mode PG Lagrangian was extended to include the appropriate pseudoscalar coupling constants that provide cross parity coupling, such terms are often referred to as ``parity violating'' terms.

The BHN Lagrangian\cite{BHN11}
has the specific form
\begin{eqnarray}
{\mathscr L}_{\hbox{\footnotesize BHN}}[\vartheta,\Gamma]&=&\frac{1}{2\kappa}\biggl[- 2 \Lambda + a_0
R + b_0X -\frac12\sum_{\mathsf{n}=1}^3 a_{\mathsf{n}}
{{}^{\mathsf{(n)}}\!T}{}^2+ 3\sigma_2
V_\mu A^\mu\biggr] \nonumber\\
&& \qquad\qquad\ \  -\frac{1}{2\varrho}\biggl[\frac{w_6}{12}R^2-\frac{w_3}{12}X^2+\frac{\mu_3}{12}RX\biggr], \label{Ldensity}
\end{eqnarray}
where
$R$ is the scalar curvature and $X$ is the
pseudoscalar curvature (specifically $X/6=R_{[0123]}$ is the
magnitude of the one independent component of the totally antisymmetric
curvature), and $V_\mu:= T^\alpha{}_{\alpha\mu}= {}^{(2)}\!
T{}^\alpha{}_{\alpha\mu}$,
$A_\mu:=\frac{1}{2}\epsilon_{\mu\nu}{}^{\alpha\beta}T^\nu{}_{\alpha\beta}
=\frac{1}{2}\epsilon_{\mu\nu}{}^{\alpha\beta} {}^{(3)}\!T{}^\nu{}_{\alpha\beta}$
are the torsion trace and axial vectors. The parameters $a_0$, $a_1$, $a_2$, $w_3$ and $w_6$ are scalars, whereas $b_0$, $\sigma_2$ and $\mu_3$ are {\em pseudoscalars\/}.
For an extensive discussion of the mathematics and physics of the PG theory and this model as well as further references see BHN\cite{BHN11} and the new work, Ref.~\refcite{BH11}.

In BHN the general field equations were worked out, and then specialized to find the most general 2-scalar mode PG FLRW cosmological model.  Here we will take an alternative approach and consider manifestly isotropic Bianchi models.

\section{The PGT scalar mode Bianchi I and IX cosmological model}
PG cosmological investigations have a long history.  For earlier PG cosmological investigations see Minkevich and coworkers, e.g., Refs.~\refcite{Min80,Min83,MN95,MG06,MGK07}
 and Goenner \& M\"uller-Hoissen;\cite{GMH}
for recent work see Refs.~\refcite{BHN11,YN07,SNY08,LSX09a,LSX09b,ALX10,WW09,JCAP09}.

For the usual FLRW models, although they are actually homogeneous and isotropic, the representation is not manifestly so.  Indeed they are merely manifestly isotropic-about-a-chosen-point.  In contrast, the representation used here for the isotropic Bianchi I and IX models has the virtue of being {\em manifestly homogeneous\/} and {\em manifestly isotropic}.  (Indeed these are the only two Bianchi models which admit such a representation.)

For homogeneous, isotropic  Bianchi type I and IX
(respectively equivalent to FLRW $k=0$ and $k=+1$) cosmological models the
isotropic orthonormal coframe has the form
\begin{equation}
\vartheta^0:=dt,\qquad \vartheta^a:=a\sigma^a,
\end{equation}
where $a=a(t)$ is the scale factor and $\sigma^j$ depends on the (not needed here) spatial coordinates in such a way that
\begin{equation}
d\sigma^i=\zeta\epsilon^i{}_{jk}\sigma^j \wedge\sigma^k,
\end{equation}
where $\zeta=0$ for Bianchi I and $\zeta=1$ for Bianchi IX, thus $\zeta^2=k$,
the sign of the FLRW Riemannian spatial curvature.

Remark: although a few other Bianchi models can be isotropic, e.g., Bianchi V, the representations themselves are not {\em manifestly isotropic}, which is an important property in our analysis below.   No negative curvature Bianchi model admits a manifestly isotropic representation.

Because of isotropy, the only non-vanishing connection one-form
coefficients are necessarily of the form
\begin{equation} \Gamma^a{}_0=\psi(t)\,
\sigma^a,\qquad \Gamma^a{}_b=\chi(t)\epsilon^a{}_{bc}\, \sigma^c.
\end{equation}
Here $\epsilon_{abc}:= \epsilon_{[abc]}$ is the usual 3 dimensional
Levi-Civita anti-symmetric symbol.

From the definition of the
curvature (\ref{curvature}), one can now find all the nonvanishing
curvature 2-form components:
\begin{eqnarray}
R^a{}_b&=&\dot \chi dt\wedge
\epsilon^a{}_{bc}\sigma^c+[\psi^2-\chi^2]\sigma^a\wedge \sigma_b
+ \chi\zeta\epsilon^a{}_{bc}\epsilon^c{}_{ij}\sigma^i\wedge\sigma^j,\\
R^a{}_0&=&\dot \psi dt\wedge \sigma^a
-\chi\psi \sigma^b\wedge\epsilon^a{}_{bc}\sigma^c
+\psi\zeta\epsilon^a{}_{bc}\sigma^b\wedge\sigma^c
.
\end{eqnarray}
Consequently, the scalar and pseudoscalar curvatures are,
respectively,
\begin{eqnarray}
R&=&6[a^{-1}\dot \psi+a^{-2}(\psi^2-[\chi-\zeta]^2+\zeta^2)], \label{R} \\
X&=&6[a^{-1}\dot \chi+2a^{-2}\psi(\chi-\zeta)].\label{X}
\end{eqnarray}

Because of isotropy, the only nonvanishing  torsion tensor
components are of the form
\begin{equation}
T^a{}_{b0}=u(t)\delta^a_b, \qquad T^a{}_{bc}=-
2x(t)\epsilon^a{}_{bc},
\end{equation}
where $u$ and $x$ are the scalar and pseudoscalar torsion, respectively.
From the definition of the torsion (\ref{torsion}), one can find the
relation between the torsion components and the gauge variables:
\begin{equation}
u=a^{-1}(\dot a-\psi), \qquad x= a^{-1}(\chi-\zeta). \label{fchi}
\end{equation}
Note that $a^{-1}\psi=H-u$, where $H=a^{-1}\dot a$ is the {\em Hubble function}.

Regarding the material source of gravity, because of the symmetry assumptions, the source material energy-momentum tensor is necessarily of the fluid form with a flow vector along the time axis and an energy density and pressure: $\rho,p$.  Although we expect the source spin density to play an
important role in the very early universe, it is reasonable to assume, as we do here,
that the material spin density at later times is negligible.

\section{Effective Lagrangian}


The dynamical equations for these homogeneous cosmologies could be obtained by imposing the Bianchi symmetry on the general field equations found by BHN from their {\em Lagrangian density}.    On the other hand dynamical equations can be obtained directly and independently (generalizing the procedure used in Ref.~\refcite{JCAP09} to include Bianchi IX, pressure and the new couplings) from a classical mechanics type {\em effective Lagrangian}, which in this case can be simply obtained by restricting the BHN Lagrangian density to the Bianchi symmetry (this step is where the manifestly homogeneous representation plays an essential role).  This procedure is known to successfully give the correct dynamical equations for all Bianchi class A models (which includes our cases) in GR,\cite{AS91} and it is conjectured to also work equally as well for the PG theory.  Our calculations explicity verify this property for the isotropic Bianchi I and IX models.  Indeed the equations we obtained in this way are equivalent to those found (at a later date) by BHN for their FLRW models (which are equivalent to our isotropic Bianchi models) by restricting to FLRW symmetry their general dynamical PG equations. This has proved to be a useful cross check.

Our {\em effective Lagrangian} $L_{\mathrm{eff}}=L_\mathrm{G}+L_{\mathrm{int}}$ includes the {\em \textit{interaction}} Lagrangian:
$L_{\mathrm{int}}= pa^3$, where $p=p(t)$ is the pressure,
and the {\em gravitational} Lagrangian:
\begin{eqnarray}
L_\mathrm{G} = \frac{1}{2\kappa}(a_0R+b_0X-2\Lambda)a^3&&+\frac3{2\kappa}(-a_2u^2+4a_3x^2+4\sigma_2ux)a^3\nonumber\\
&&-\frac1{24\varrho}(w_6R^2-w_3X^2+\mu_3RX)a^3.
\end{eqnarray}
It should be noted that the parameter restrictions $a_2<0$, $w_6<0$, $w_3>0$, and $\mu_3^2+4w_3w_6<0$
are, in the light of eqs. (\ref{R}), (\ref{X}), (\ref{fchi}), {\em necessary} for the {\em least action principle}, which requires {\em positive\/} quadratic-kinetic-terms.

In the following we often take for simplicity units such that $\kappa=1=\varrho$.  These factors can be easily restored in the final results by noting that in the Lagrangian they occur in conjunction with certain PG parameters.  Hence in the final results one need merely make the replacements
$\{a_0,a_2,a_3,b_0,\Lambda,\sigma_2\}\to \kappa^{-1}\{a_0,a_2,a_3,b_0,\Lambda,\sigma_2\}$, $\{w_3,w_6,\mu_3\}\to \varrho^{-1}\{w_3,w_6,\mu_3\}$.

The gravitational Lagrangian has the usual form of a sum of terms homogeneous in ``velocities'' $L_\mathrm{G}=L_0+L_1+L_2$; the associated {\em energy function} is thus
\begin{eqnarray}
{\cal E}_\mathrm{G}&:=&\frac{\partial L_\mathrm{G}}{\partial \dot \psi}\dot\psi+\frac{\partial L_\mathrm{G}}{\partial \dot\chi}\dot\chi+\frac{\partial L_\mathrm{G}}{\partial \dot a}\dot a-L_\mathrm{G}=L_2-L_0\nonumber\\
&=&a^3\Biggl\{-3(a_0-\frac12a_2)u^2-3a_0H^2+3x^2(a_0-2a_3)\nonumber\\
&&+6uH(a_0-\frac12a_2)+6(b_0+\sigma_2)x(H-u)-3a_0\frac{\zeta^2}{a^2}+\Lambda\nonumber\\
&&-\frac{w_6}{24}\left[R^2-12R\left\{(H-u)^2-x^2+\frac{\zeta^2}{a^2}\right\}\right]\nonumber\\
&&+\frac{w_3}{24}\left[X^2+24Xx(H-u)\right]\nonumber\\
&&-\frac{\mu_3}{24}\left[RX-6X\left\{(H-u)^2-x^2+\frac{\zeta^2}{a^2}\right\}+12Rx(H-u)\right]\Biggr\}.\label{energyfunction}
\end{eqnarray}
The energy value (\ref{energyfunction}) has the form $-a^3\rho$ where $\rho$ can be identified as the material {\em energy density}.

Making use of the formulas for the torsion and curvature components
in terms of the gauge variables (\ref{R},\ref{X},\ref{fchi}), we now
obtain the Euler-Lagrange equations $\displaystyle{ \frac{d}{d t}
\frac{\partial L_{\mathrm{eff}}}{\partial \dot{q_k}} -
\frac{\partial L_{\mathrm{eff}}}{\partial q_k} =0}$,
where $q_k=\{\psi,\chi, a\}$. The dynamical equations are the $\psi$ equation:
\begin{eqnarray}
\frac{d}{dt}\frac{\partial L_\mathrm{G}}{\partial \dot
\psi}&=&\frac{d}{dt}\left(a^{2}\left[3a_0-\frac{w_6}2R-\frac{\mu_3}4X\right]\right)=\frac{\partial L_\mathrm{G}}{\partial \psi}\nonumber\\
&=&3(a_2u-2\sigma_2x)a^2 +\left[6a_0-w_6R-\frac{\mu_3}2X
\right]a\psi   \nonumber \\
&&+\left[6b_0-\frac{\mu_3}2 R+w_3X\right]a(\chi-\zeta),
\end{eqnarray}
which can be rearranged into the form
\begin{equation}
-\frac{w_6}2\dot R-\frac{\mu_3}4 \dot
X=-\left[3(2a_0-a_2)-w_6R-\frac{\mu_3}2
X\right]u-\left[6(b_0+\sigma_2)-\frac{\mu_3}2 R+w_3X\right]x; \label{Epsi}
\end{equation}
the $\chi$ equation:
\begin{eqnarray}
\frac{d}{dt}\frac{\partial L_\mathrm{G}}{\partial \dot
\chi}&=&\frac{d}{dt}\left(a^{2}\left[3b_0-\frac{\mu_3}4R+\frac{w_3}2X\right]\right)=\frac{\partial L_\mathrm{G}}{\partial \chi}\nonumber\\
&=&-6(2a_3x+\sigma_2u)a^2 -\left[6a_0-w_6R-\frac{\mu_3}2
X\right]a(\chi-\zeta) \nonumber \\
& &+\left[6b_0-\frac{\mu_3}2 R+w_3X\right]a\psi,
\end{eqnarray}
which can be rearranged into the form
\begin{equation}
-\frac{\mu_3}4 \dot
R+\frac{w_3}2\dot X=-\left[6(b_0+\sigma_2)-\frac{\mu_3}2 R+w_3X\right]u+\left[6(a_0-2a_3)-w_6R-\frac{\mu_3}2
X\right]x; \label{Echi}
\end{equation}
and the $a$ equation:
\begin{eqnarray}
\frac{d}{dt}\frac{\partial L_\mathrm{G}}{\partial \dot
a}&=&\frac{d}{dt}\left(-a^{2}3[a_2u-2\sigma_2x]\right)=\frac{\partial L_\mathrm{G}}{\partial a}
      +\frac{\partial L_\mathrm{int}}{\partial a}\\
&=&3a^{-1}L-\left(\frac{a_0}2-\frac{w_6}{12}R-\frac{\mu_3}{24}X\right)[a^2R+6(\psi^2-[\chi-\zeta]^2+\zeta^2)]\nonumber\\
&&-\left(\frac{b_0}2+\frac{w_3}{12}X-\frac{\mu_3}{24}R\right)[a^2X+12\psi(\chi-\zeta)]\nonumber\\
&&+3a^2(a_2u-2\sigma_2x)u-6a^2[2a_3x+\sigma_2u]x+3pa^2,
\end{eqnarray}
which can be rearranged into the form
\begin{eqnarray}
-3(a_2\dot u-2\sigma_2 \dot x)&=&6H[a_2u-2\sigma_2x]+
\frac32(a_0R+b_0X-2\Lambda)\nonumber\\
&&+\frac9{2}(-a_2u^2+4a_3x^2+4\sigma_2ux)
-\frac1{8}(w_6R^2-w_3X^2+\mu_3RX)\nonumber
\\
&&+\left(\frac{a_0}2-\frac{w_6}{12}R-\frac{\mu_3}{24}X\right)\left[-R-6(H-u)^2
+6x^2-6\frac{\zeta^2}{a^2}\right]\nonumber\\
&&+\left(\frac{b_0}2+\frac{w_3}{12}X-\frac{\mu_3}{24}R\right)\left[-X+12x(H-u)\right]\nonumber\\
&&+3(a_2u-2\sigma_2x)u-6[2a_3x+\sigma_2u]x+3p\\ \Longrightarrow\hspace{1.7cm}&& \nonumber\\
-3(a_2\dot u-2\sigma_2 \dot x) &=&
(a_0R+b_0X-3\Lambda)+6H[a_2u-2\sigma_2x]\nonumber\\
&&+\frac3{2}(-a_2u^2+4a_3x^2+4\sigma_2ux)
-\frac1{24}(w_6R^2-w_3X^2+\mu_3RX)\nonumber
\\
&&+\frac{a_0}2\left[-6(H-u)^2+6x^2-6\frac{\zeta^2}{a^2}\right]
+6b_0x(H-u)\nonumber
\\
&&-\frac1{24}\left(2w_6R+\mu_3X\right)\left[-6(H-u)^2+6x^2
-6\frac{\zeta^2}{a^2}\right]\nonumber\\
&&+\frac1{24}\left(2w_3X-\mu_3R\right)\left[12x(H-u)\right]+3p. \label{Ea}
\end{eqnarray}

\bigskip
Since $L_{\rm G}$ is time independent, the energy function satisfies an {\em energy conservation} relation:
\begin{equation}
\dot{\cal E}=-\frac{\delta L_{\rm G}}{\delta q^k} {\dot q}{}^k 
=-\frac{\delta L_{\rm G}}{\delta \psi} {\dot \psi}-\frac{\delta L_{\rm G}}{\delta \chi} {\dot \chi}-\frac{\delta L_{\rm G}}{\delta a} {\dot a}=\frac{\delta L_{\rm int}}{\delta a}{\dot a}=3pa^2{\dot a},
\end{equation}
hence, as expected, we recover the perfect fluid relation
\begin{equation} -\frac{d(\rho a^3)}{dt}=p\frac{da^3}{dt}.\end{equation}

The above equations (\ref{Epsi},\ref{Echi},\ref{Ea}) are 3 {\em second order} equations for the gauge potentials $a,\psi,\chi$.
However they can in an alternative way be used as part of a set of 6 {\em first order} equations along with the Hubble relation
$
\dot a=aH
$ 
and the following two relations, obtained by taking the time
derivatives of the torsion (\ref{fchi}) and using the curvature definitions (\ref{R},\ref{X}):
\begin{eqnarray}
\qquad\dot x&=&-Hx-\frac{X}6-2x(H-u),\\
\dot H -\dot u&=&\frac{W}6-H(H-u)-(H-u)^2+x^2-\frac{\zeta^2}{a^2}.
\end{eqnarray}
One advantage of such a reformulation is that the variables are now all observables.

Our 6 first order dynamical equations and the energy constraint equation can now be put in the form
\begin{eqnarray}
\qquad\qquad\dot a&=&aH,\label{dota}\\
\qquad\qquad\dot H&=&
  \frac{1}{6a_2}( \tilde a_2R-2\tilde \sigma_2 X)-2H^2
  +\frac{\tilde a_2-4\tilde a_3}{a_2}x^2-\frac{\zeta^2}{a^2}\nonumber\\
&&+\frac{(\rho-3p)}{3a_2}+\frac{4\Lambda}{3a_2}\label{dotH},\\
\qquad\qquad\dot u&=&
  -\frac1{3a_2}(a_0R+\tilde \sigma_2 X)-3Hu
  +u^2-\frac{4a_3}{a_2}x^2
  +\frac{(\rho-3p)}{3a_2}+\frac{4\Lambda}{3a_2},\qquad\label{dotu} \\
\qquad\qquad\dot x&=&-\frac{X}6-(3H-2u)x\label{dotx},\\
\quad-\frac{w_6}2\dot R-\frac{\mu_3}4 \dot X&=&
  \left[3\tilde a_2+w_6R+\frac{\mu_3}2
X\right]u+\left[-6\tilde \sigma_2+\frac{\mu_3}2 R-{w_3}X\right]x\label{dotRdotX}\\
\quad\,\,\,\frac{w_3}2\dot X-\frac{\mu_3}4 \dot R &=&
  \left[-6\tilde \sigma_2+\frac{\mu_3}2 R-w_3X\right]u
  -\left[12\tilde a_3+{w_6}R+\frac{\mu_3}2X\right]x\label{dotXdotR},
\end{eqnarray}
\begin{eqnarray}
\rho&=&
         3(-\frac12 \tilde a_2+2\tilde a_3)x^2+\frac{3a_2}{2}\left[H^2
       +\frac{\zeta^2}{a^2}\right]-\Lambda\nonumber\\
  &&+\left(-6\tilde \sigma_2+\frac{\mu_3}{2}R-w_3X\right)x(H-u)
       +\frac{1}{24}(w_6R^2-w_3X^2+\mu_3RX)\nonumber\\
  &&-\frac12\left(3 \tilde a_2+w_6R+\frac{\mu_3}{2}X\right)
       \left[(H-u)^2-x^2+\frac{\zeta^2}{a^2}\right], \label{rho}
\end{eqnarray}
where we have introduced certain {\em modified} parameters
$\tilde{a}_2$, $\tilde{a}_3$ and $\tilde{\sigma}_2$
with the definitions
\begin{equation}\label{mp}
    \tilde{a}_2:=a_2-2a_0, \qquad\tilde{a}_3:=a_3-\frac{a_0}{2}, \qquad\tilde{\sigma}_2:=\sigma_2+b_0.
\end{equation}
The last two dynamical equations (\ref{dotRdotX},\ref{dotXdotR})
can of course be resolved for $\dot R$ and $\dot X$:
\begin{eqnarray}
\dot R &=&
  \frac{6}{\alpha}\left[(w_3 \tilde a_2-\mu_3\tilde \sigma_2)u
                -(2w_3\tilde \sigma_2+2\mu_3\tilde a_3)x\right]\nonumber\\
                &&-2Ru-\frac{(4w_3^2+\mu_3^2)}{2\alpha}Xx
                +\frac{(w_3-w_6)\mu_3}{\alpha}Rx,\\
\dot X&=&
  \frac{6}{\alpha}\left[(2w_6\tilde \sigma_2 +\frac12 \mu_3\tilde a_2)u
             +(4w_6\tilde a_3-\mu_3\tilde \sigma_2)x\right]\nonumber\\
             &&-2Xu+\frac{(4w_6^2+\mu_3^2)}{2\alpha}Rx
             -\frac{(w_3-w_6)\mu_3}{\alpha}Xx,\label{N-mode}
\end{eqnarray}
where
\begin{equation}\label{alpha}
    \alpha:=-w_3w_6-\frac{\mu_3^2}{4}.
\end{equation}
Note that for the range of parameters of physical interest $\alpha>0$.

We have cast our system into six first order equations for (3D)
tensorial quantities, equations which are suitable for numerical
evolution and comparison with observations. However these equations
are probably not in the most suitable form for the most penetrating
analytic analysis. So we here also present the  Hamiltonian equations
for our PG cosmology.

\section{Hamiltonian formulation}

From the above one can introduce the canonical conjugate momentum
variables:
\begin{eqnarray}
P_a
&\equiv&\frac{\partial L_{\mathrm{eff}}}{\partial \dot a}=-3a^2\left[a_2u-2\sigma_2x\right],\\
P_\psi
&\equiv&\frac{\partial L_{\mathrm{eff}}}{\partial \dot \psi}=a^2\left[3a_0-\frac{w_6}{2}R-\frac{\mu_3}{4}X\right], \label{ppsi}\\
P_\chi
&\equiv&\frac{\partial L_{\mathrm{eff}}}{\partial \dot \chi}=a^2\left[3b_0+\frac{w_3}{2}X-\frac{\mu_3}{4}R\right]. \label{pchi}
\end{eqnarray}
Now one can construct the {\em effective Hamiltonian}:
\begin{eqnarray}
\qquad\qquad\mathcal{H}_{\mathrm{eff}}&=&P_a\dot a+P_\psi\dot\psi+P_\chi\dot\chi-L_\mathrm{eff}\nonumber\\
&=&a^3(\Lambda-p)-6aa_3(\chi-\zeta)^2+\frac{3\sigma_2^2a^2(\chi-\zeta)}{a_2}\nonumber\\
&&
      -\frac{3a^3}{2\alpha}(w_3a_0^2-w_6b_0^2+\mu_3a_0b_0) \nonumber\\
&& +P_a\left[\frac{\sigma_2}{a_2}\left(\frac a2-\chi+\zeta\right)+\psi\right] \nonumber\\
&& +P_\psi\left[-\psi^2+(\chi-\zeta)^2-\zeta^2+\frac{(b_0\mu_3-2a_0w_3)a^2}{2\alpha}\right]\frac 1{a} \nonumber\\
&& +P_\chi\left[-2\psi(\chi-\zeta)+\frac{(a_0\mu_3+2b_0w_6)a^2}{2\alpha}\right]\frac1{a} \nonumber\\
&& +P_\psi P_\chi\left[-\frac{\mu_3}{6\alpha}\right]\frac 1{a}
     +P_\psi^2\left[-\frac{w_3}{6\alpha}\right]\frac 1{a}
     +P_\chi^2\left[\frac{w_6}{6\alpha}\right]\frac 1{a}
     +P_a^2\left[-\frac 1{6a_2}\right]\frac 1{a}.\qquad
\end{eqnarray}

From the effective Hamiltonian, we obtain
the six first order Hamilton equations:
\begin{eqnarray}
\dot a &=&\frac{\partial \mathcal H_{\mathrm{eff}}}{\partial P_a}
   =\left[\frac{\sigma_2}{a_2}\left(\frac a2-\chi+\zeta\right)+\psi\right]-\frac{P_a}{3a_2a}, \\
\dot \psi &=&\frac{\partial \mathcal H_{\mathrm{eff}}}{\partial P_\psi}
   =\frac 1{a}\left[-\psi^2+(\chi-\zeta)^2-\zeta^2-\frac{\mu_3(3a^2b_0-P_\chi)-2w_3(3a^2a_0+P_\psi)}{6\alpha}\right],\\
\dot \chi &=&\frac{\partial \mathcal H_{\mathrm{eff}}}{\partial P_\chi}
   =\frac1{a}\left[-2\psi(\chi-\zeta)-\frac{\mu_3(3a^2a_0-P_\psi)+2w_6(3a^2b_0+P_\chi)}{6\alpha}\right],\\
\dot{P_a} &=&-\frac{\partial \mathcal H_{\mathrm{eff}}}{\partial a}
   =a^{-1}\mathcal{H}_{\mathrm{eff}}-a^{-1}P_a\left[\frac{\sigma_2}{a_2}(a-\chi+\zeta)+\psi\right]\nonumber\\
   &&
    -4a^2\left[\frac{3(w_3a_0^2-w_6b_0^2+\mu_3a_0b_0)}{2\alpha}-(\Lambda-p)\right] \nonumber\\
          & & +12a_3(\chi-\zeta)^2+P_\psi\frac{(b_0\mu_3-2a_0w_3)}{2\alpha}
    +P_\chi\frac{(a_0\mu_3+2b_0w_6)}{2\alpha}-\frac{9\sigma_2^2a(\chi-\zeta)}{a_2},\qquad \\
\dot{P_\psi} &=&-\frac{\partial \mathcal H_{\mathrm{eff}}}{\partial \psi}
   =-P_a+\frac 2a\left[P_\psi\psi+P_\chi(\chi-\zeta)\right], \\
\dot{P_\chi} &=&-\frac{\partial \mathcal H_{\mathrm{eff}}}{\partial \chi}
   =12aa_3\chi-\frac{3\sigma_2^2a^2}{a_2}+P_a\frac{\sigma_2}{a_2}+\frac 2a[P_\chi\psi-P_\psi(\chi-\zeta)].
\end{eqnarray}

This canonical reformulation should be of considerable interest for
further studies of this model, since the Hamiltonian formulation is
the framework for the most powerful known approaches for
analytically studying the dynamics of a system, including such
techniques as the Hamilton-Jacobi method and phase space portraits.

\section{Numerical Demonstration}
\begin{figure}[thbp]
\begin{tabular}{rl}
\includegraphics[width=0.482\textwidth]{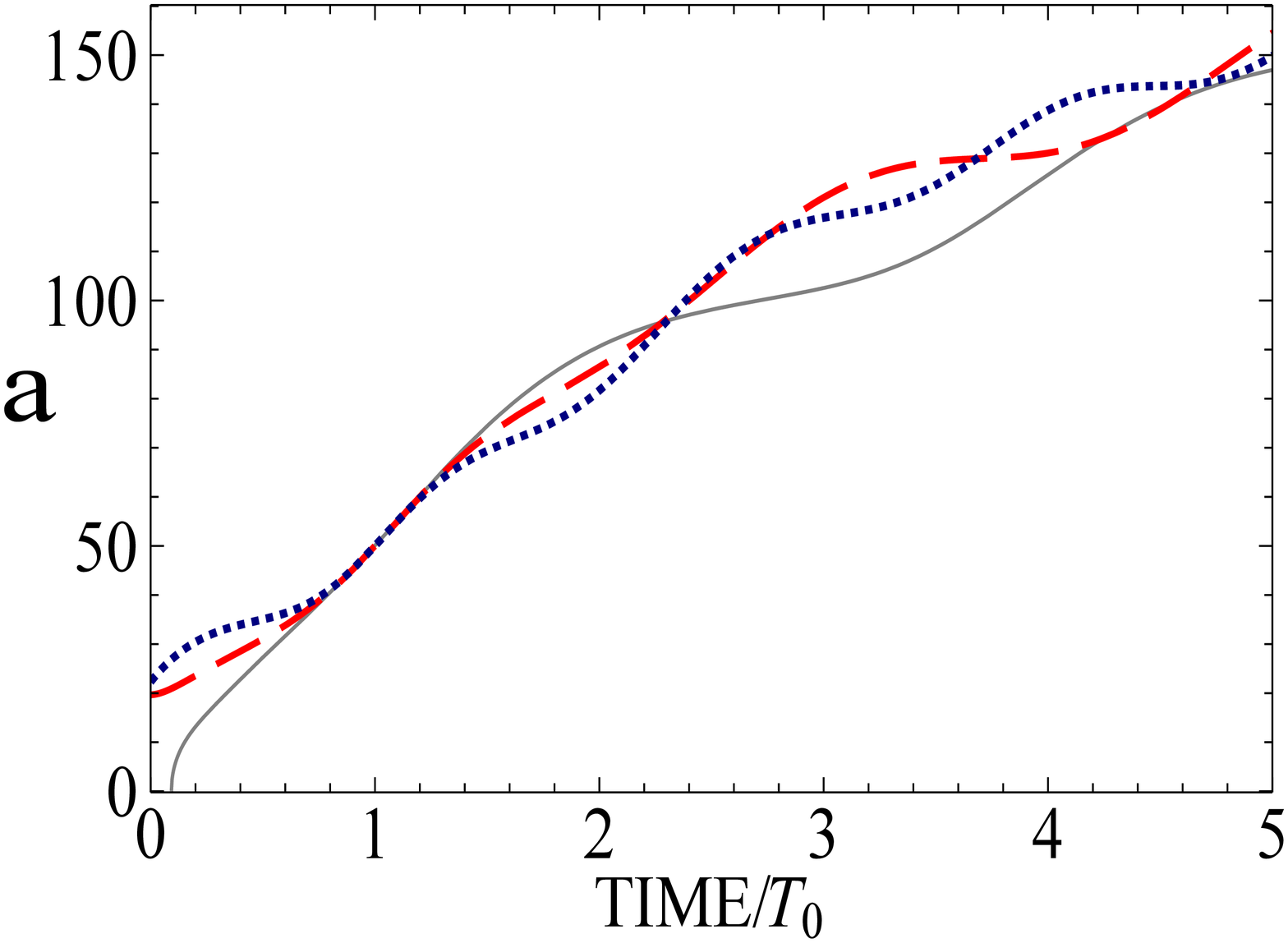}&
\includegraphics[width=0.482\textwidth]{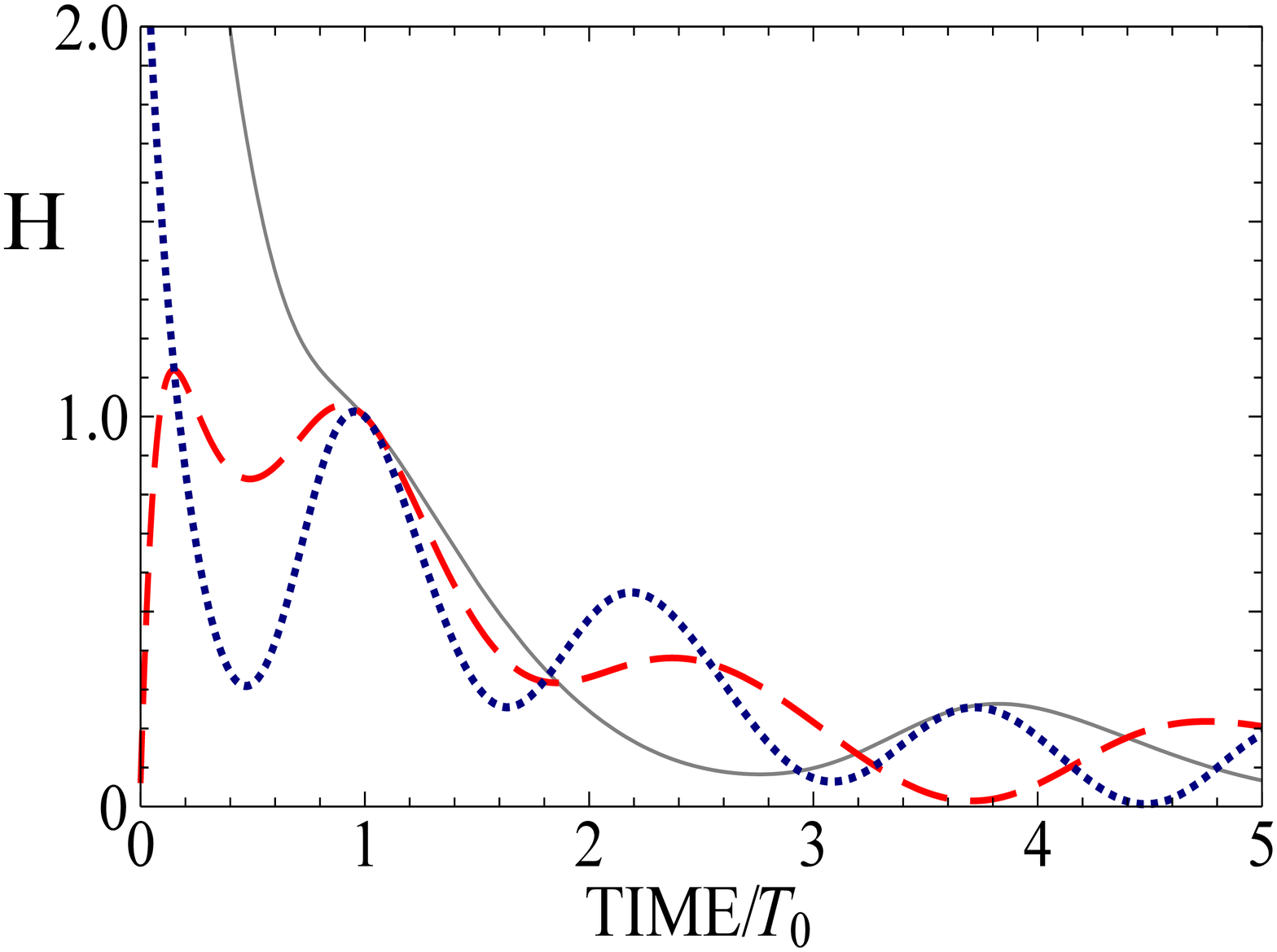}\\
\includegraphics[width=0.482\textwidth]{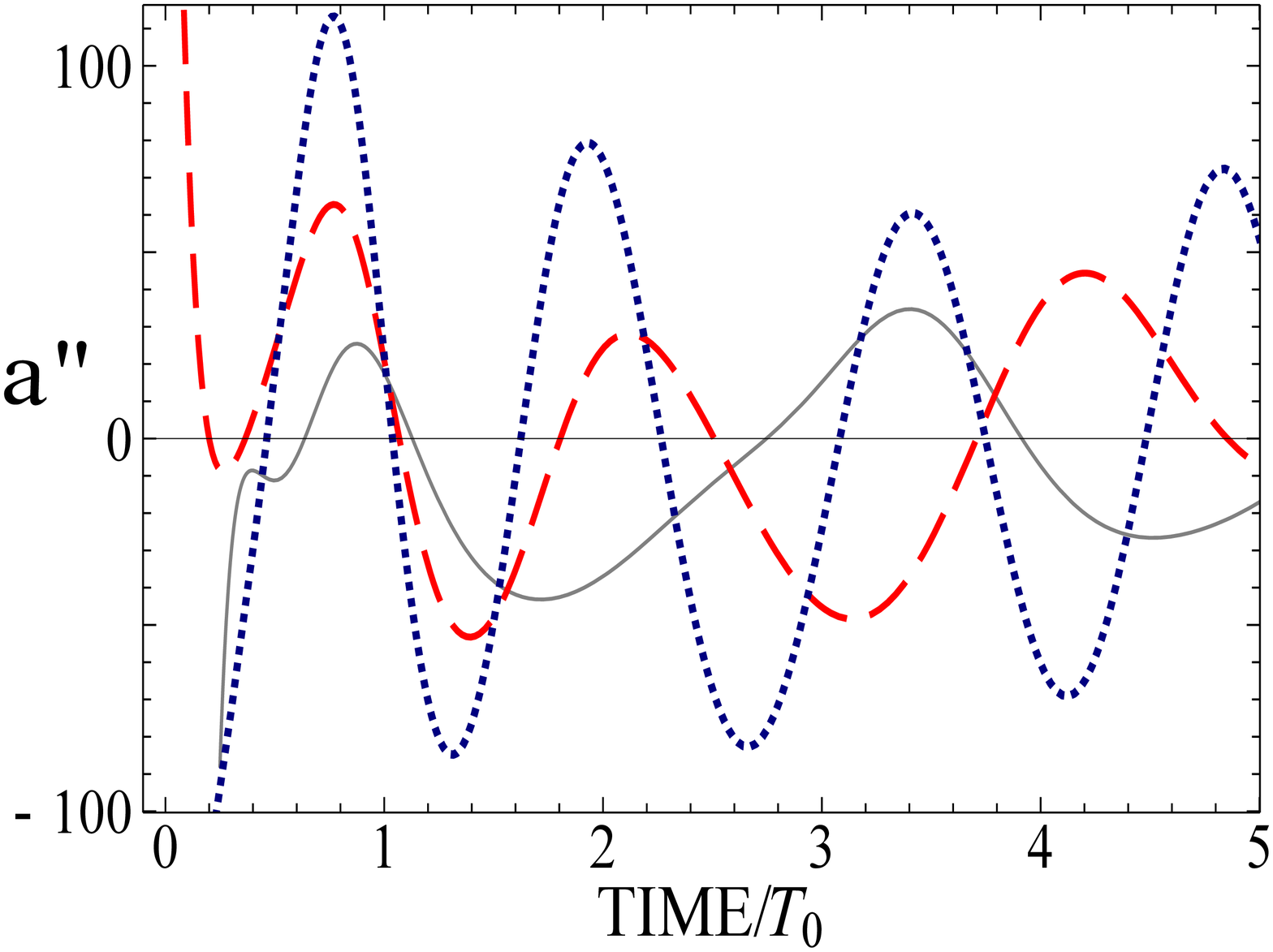}&
\includegraphics[width=0.482\textwidth]{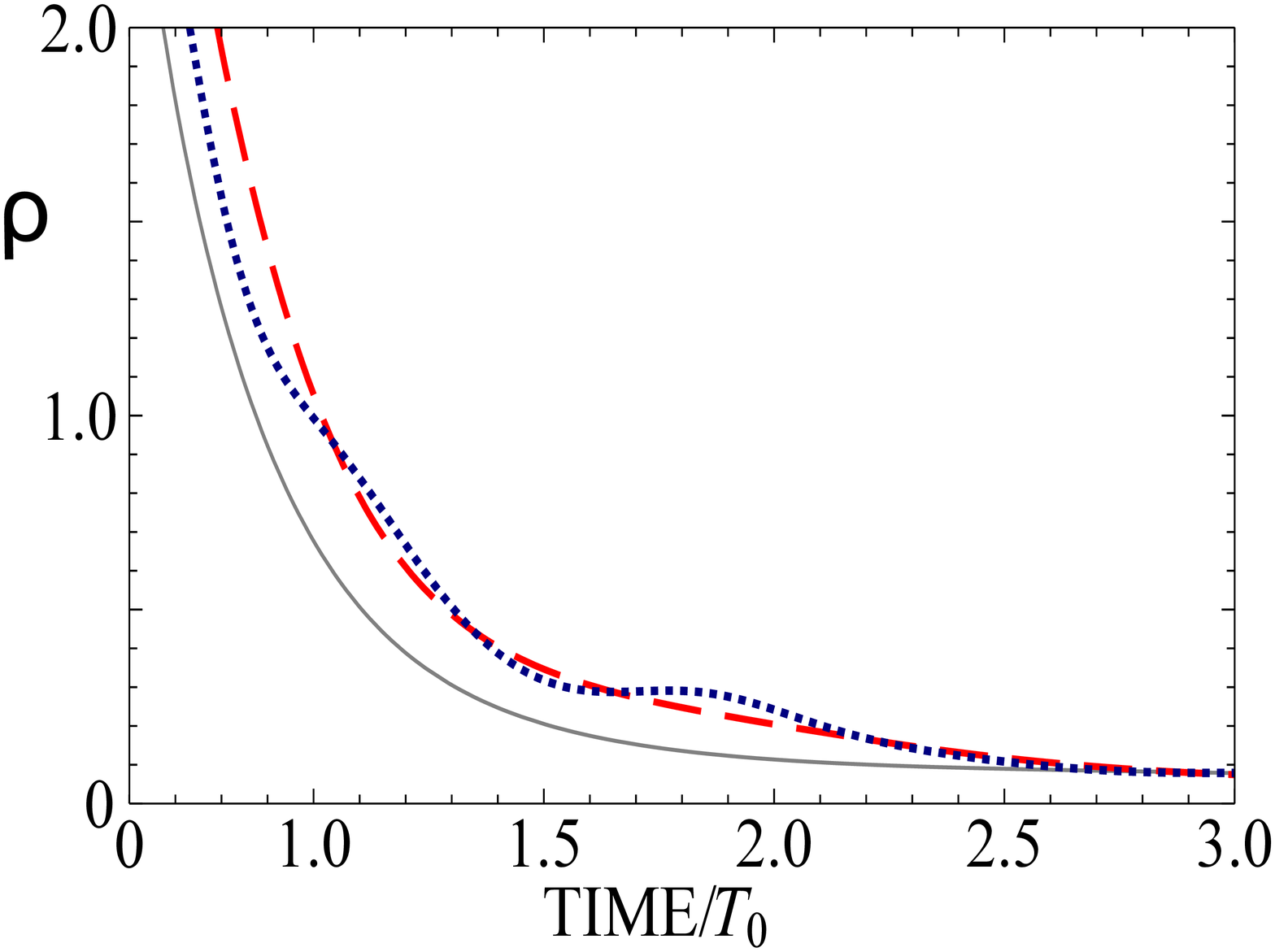}\\
\end{tabular}
\caption{Evolution of the expansion factor $a$, the Hubble function,
$H$, the 2nd time derivative of the expansion factor, $\ddot a$
and the mass density, $\rho$ as functions of time with the parameter choice and the initial data as specified in the text.
 The gray (solid) line represents the evolution with the pseudoscalar parameters $\sigma_2$ and $\mu_3$ turned off.
The red (dashed) line represents the evolution with the parameter $\sigma_2$ activated.
The blue (doted) line represents the evolution including both pseudoscalar parameters $\sigma_2$ and $\mu_3$.} \label{zA1}
\end{figure}

\begin{figure}[thbp]
\begin{tabular}{ccc}
\includegraphics[width=0.307\textwidth]{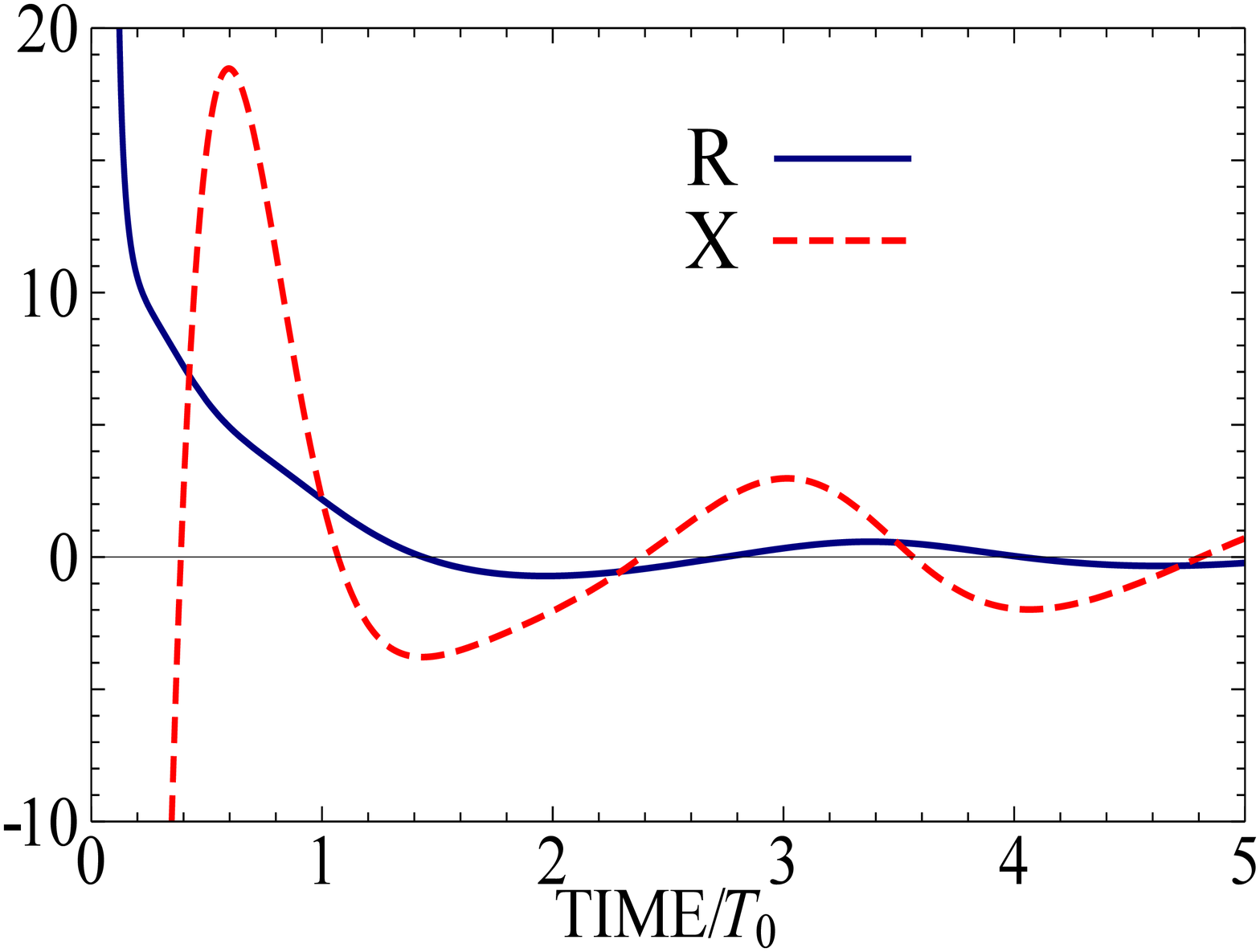}&
\includegraphics[width=0.307\textwidth]{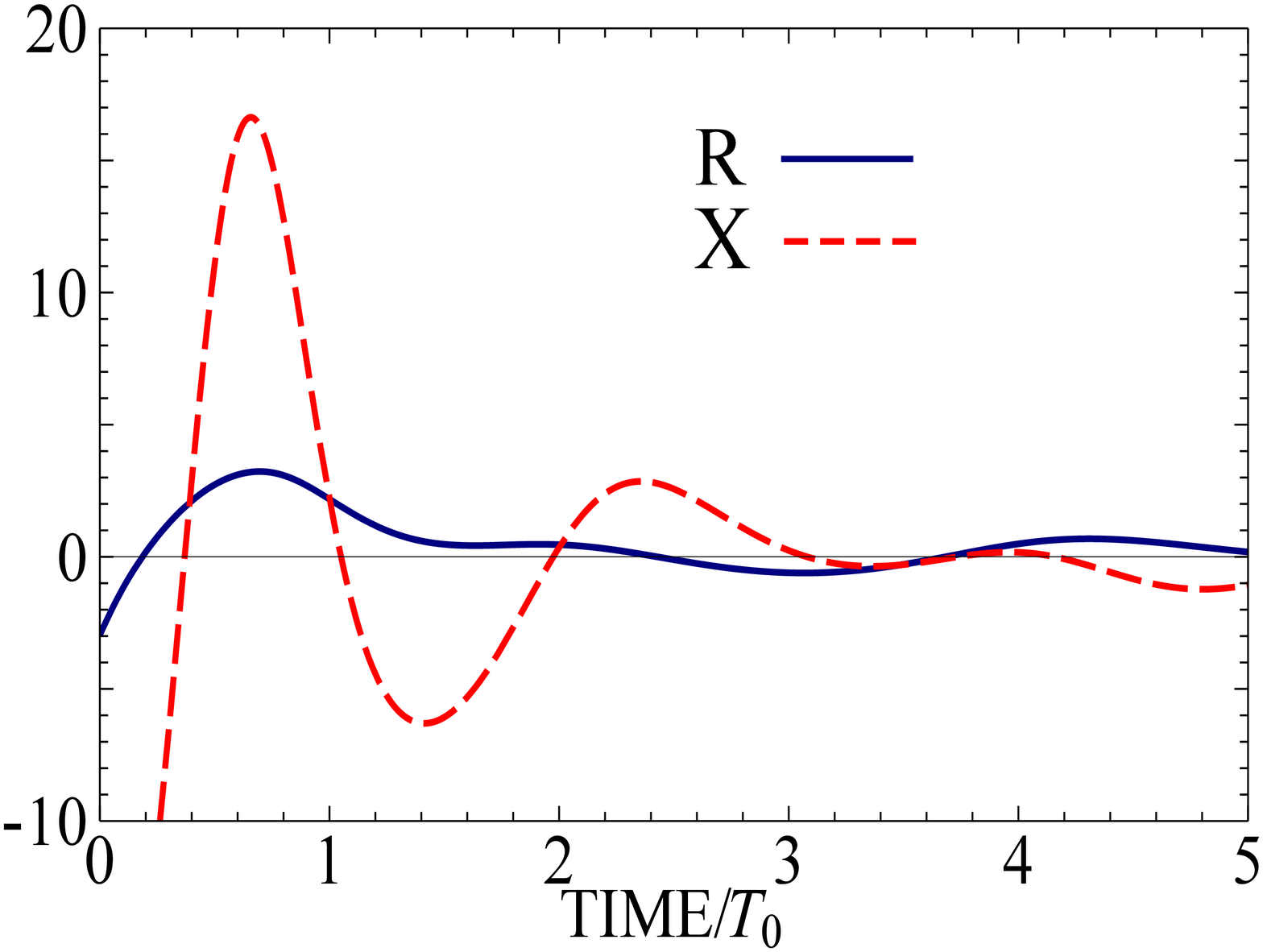}&
\includegraphics[width=0.307\textwidth]{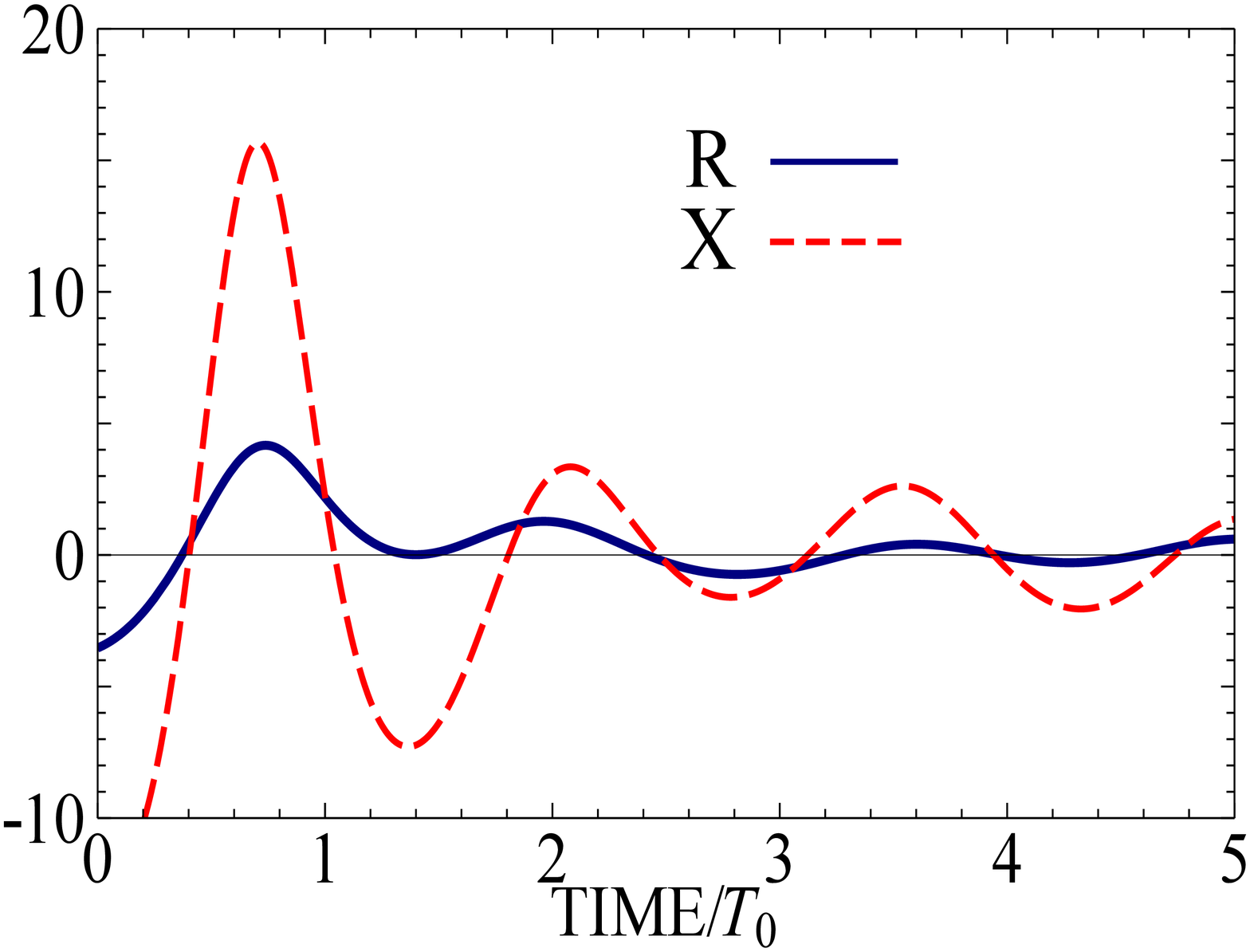}\\
\includegraphics[width=0.307\textwidth]{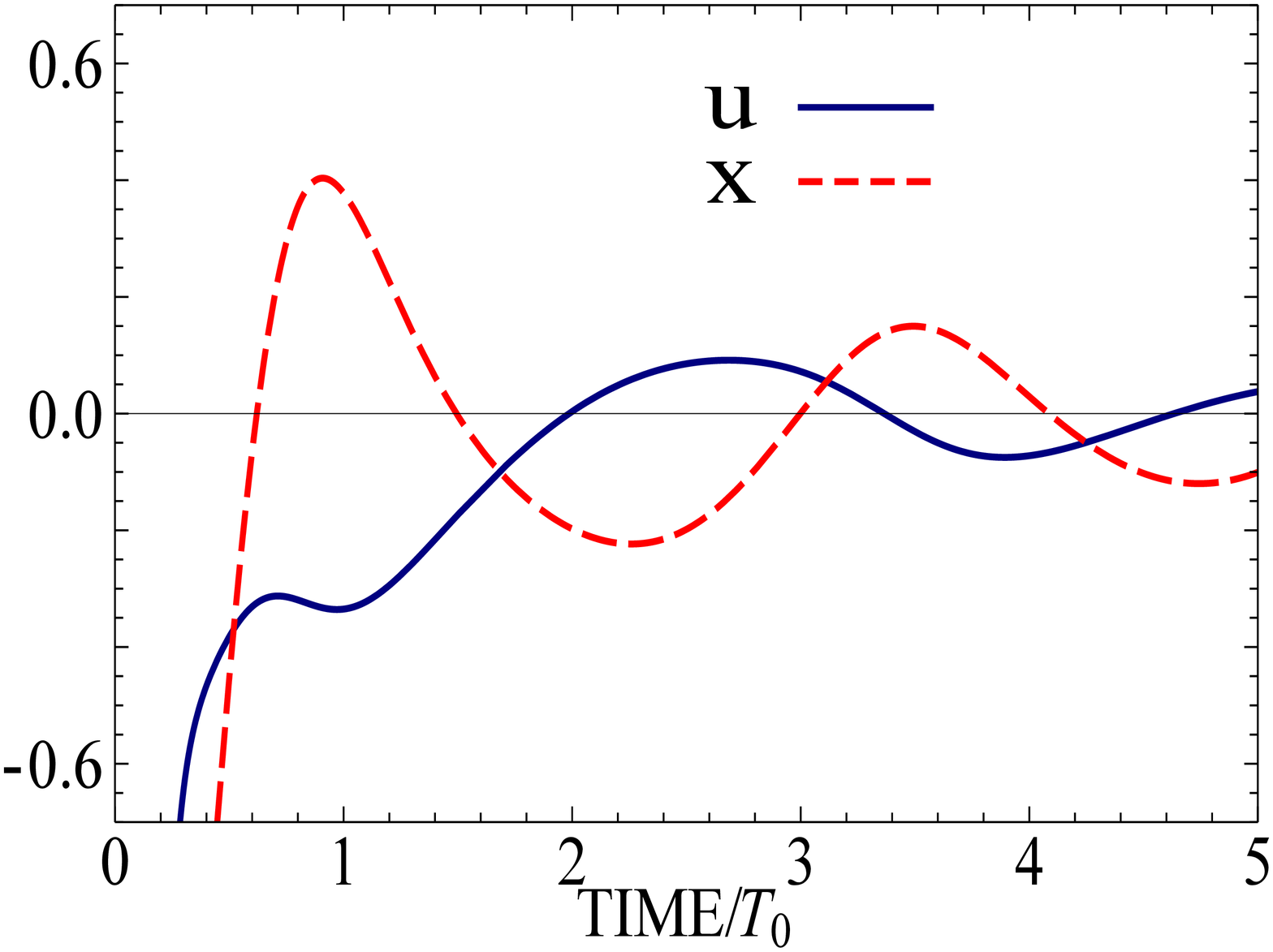}&
\includegraphics[width=0.307\textwidth]{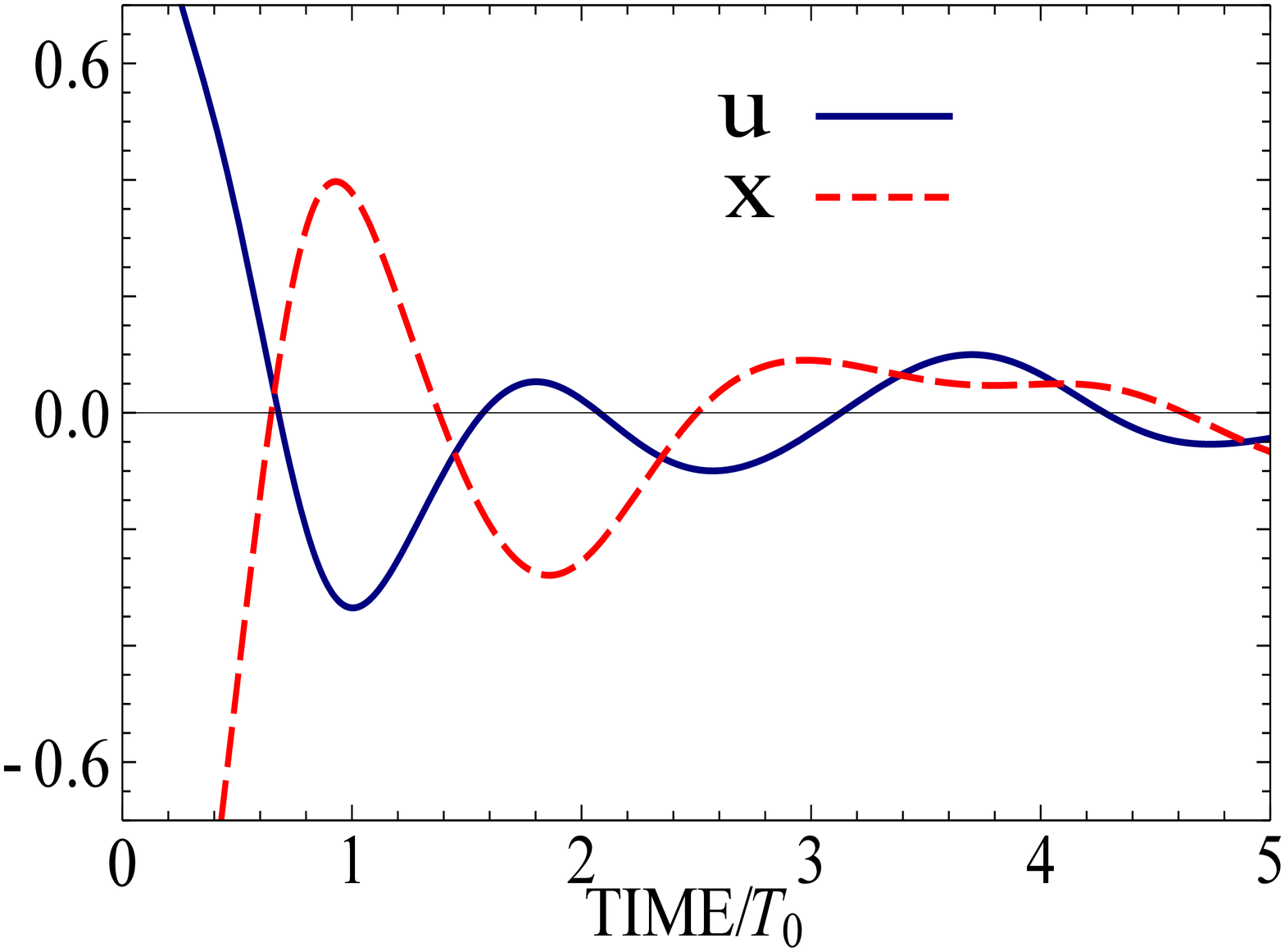}&
\includegraphics[width=0.307\textwidth]{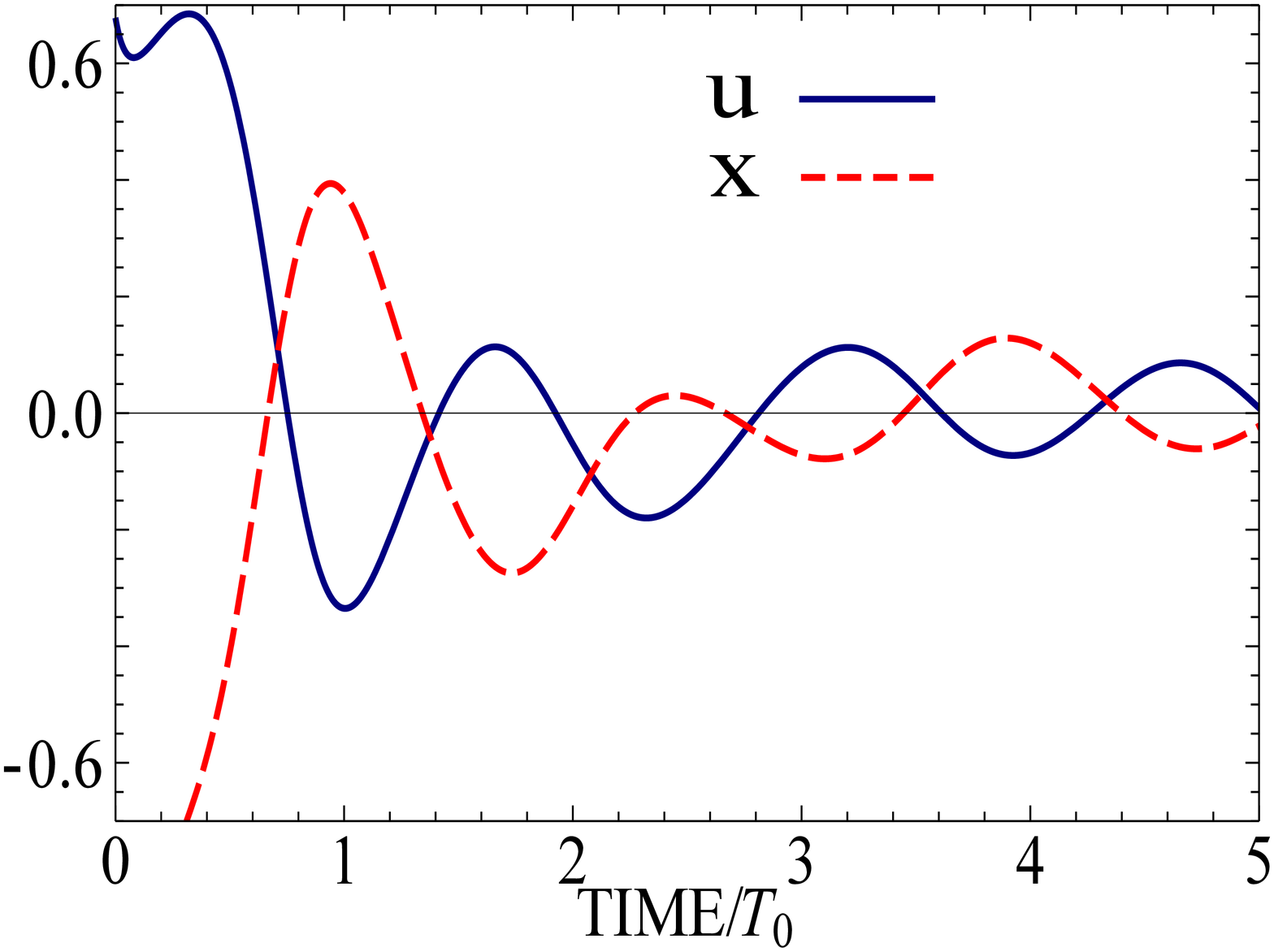}\\
\end{tabular}
\caption{In the first line we compare the scalar curvature, $R$
and the pseudoscalar curvature, $X$ in different situations.
In the second line we compare the torsion, $u$ and the axial torsion, $x$.
The first column is the evolution with vanishing pseudoscalar parameters, $\sigma_2$ and $\mu_3$,
the second column, with parameter $\sigma_2$, the third column, with both pseudoscalar parameters, $\sigma_2$ and $\mu_3$.
} \label{zA2}
\end{figure}

We present the results of a numerical evolution of
our cosmological model. For all these calculations we take
$\Lambda=0$ and $p=0$. We need to look into the scaling features of this model before we
can obtain the sort of evolution results we seek on a cosmological
scale. In terms of fundamental units we can scale the variables and
the parameters as $\kappa=8\pi G=1$. So the variables and the scaled
parameters $w_6$ and $w_3$ become dimensionless (note: the Newtonian
limit gives $a_0=1$). However, as we are interested in the cosmological
scale to see changes on the order of the age of our Universe, let us
introduce a dimensionless constant $T_0$, which represents the
magnitude of the Hubble time ($T_0=H^{-1}_0\doteq4.41504\times
10^{17}$ seconds). With this scaling, all the field equations are kept unchanged while
the period $T\rightarrow T_0T$.

For our example evolution we take the parameters as
\begin{eqnarray}
a_0&=&1,\quad a_2=-0.83,\quad a_3=-0.35,\quad w_{6}=-1.1,\quad w_{3}=0.091,\nonumber\\
\sigma_2&=&0.26 \quad \hbox{and} \quad \mu_3=0.21.
\end{eqnarray}
The behavior of the 6 equations has been observed with several
sets of initial values. We plot this typical case in two sets of figures.
In Fig.~\ref{zA1}: the evolution of the expansion factor a, the Hubble function, $H$,
 the second time derivative of the expansion factor, $\ddot{a}$,
and the energy density, $\rho$.
In Fig.~\ref{zA2}: first line, the scalar curvature and the pseudoscalar curvature, $R$ and $X$,
second line, the torsion and the axial torsion function, $u$ and $x$.

To show the effect of the new pseudoscalar coupling parameters $\tilde\sigma_2$, $\mu_3$ we have shown the evolution with these parameters turned off, with only $\tilde\sigma_2$, and with both pseudoscalar parameters activated.


\section{Concluding discussion}

 We have been investigating the dynamics of the Poincar\'e gauge theory of gravity.
  Recently, the model  with two good propagating modes carrying spin $0^+$, $0^-$
  (referred to as the scalar and pseudoscalar modes) has been extended to
  include pseudoscalar constants that couple the two different parity modes.
  Here we have considered the dynamics of this BHN model in the context of
  manifestly homogeneous and isotropic cosmological models.  We found an effective Lagrangian
 and Hamiltonian and the associated dynamical equations.  The Lagrangian equations were rearranged into a
  system  of 6 first order equations suitable for numerical evolution and a sample evolution was presented which showed the effect of the pseudoscalar coupling constants---which provide a direct interaction between the even and odd parity modes.

In these models, at late times the acceleration oscillates.  It can be positive
at the present time.
It should be noted that the $0^+$ torsion does not directly couple to any known form of matter, but it does couple directly to the Hubble expansion, and thus can directly influence the acceleration of the universe.
On the other hand, the $0^-$ couples directly to fundamental fermions; with the newly introduced pseudoscalar coupling constants it too can directly influence the cosmic acceleration.

The objective of the present work was to derive certain dynamical equations for the PG BHN isotropic homogeneous cosmological model. These will serve as a foundation for our future investigations into the dynamics of this model.

\section*{Acknowledgments}
This work was supported by the National Science Council of the R.O.C. under the grants NSC-98-2112-M-008-008 and NSC-99-2112-M-008-004 and in part by the National Center of Theoretical Sciences (NCTS).


\end{document}